\documentclass[5p,twocolumn]{elsarticle}
\usepackage{amssymb}

\setcitestyle{open={[},close={]}} %Citation-related commands

\journal{Nuclear Instruments and Methods in Physics Research Section A}

\begin{document}

\begin{frontmatter}

%% Title, authors and addresses

%% use the tnoteref command within \title for footnotes;
%% use the tnotetext command for theassociated footnote;
%% use the fnref command within \author or \affiliation for footnotes;
%% use the fntext command for theassociated footnote;
%% use the corref command within \author for corresponding author footnotes;
%% use the cortext command for theassociated footnote;
%% use the ead command for the email address,
%% and the form \ead[url] for the home page:
%% \title{Title\tnoteref{label1}}
%% \tnotetext[label1]{}
%% \author{Name\corref{cor1}\fnref{label2}}
%% \ead{email address}
%% \ead[url]{home page}
%% \fntext[label2]{}
%% \cortext[cor1]{}
%% \affiliation{organization={},
%%            addressline={}, 
%%            city={},
%%            postcode={}, 
%%            state={},
%%            country={}}
%% \fntext[label3]{}

\title{Recovery of HADES drift chambers suffering from Malter-like effects}

%% use optional labels to link authors explicitly to addresses:
%% \author[label1,label2]{}
%% \affiliation[label1]{organization={},
%%             addressline={},
%%             city={},
%%             postcode={},
%%             state={},
%%             country={}}
%%
%% \affiliation[label2]{organization={},
%%             addressline={},
%%             city={},
%%             postcode={},
%%             state={},
%%             country={}}

%%\author{Christian Wendisch, Luis Alberto Vieira Lopes, Christian Müntz}

\author[a,1]{Christian Wendisch}
\author[b]{Christian Müntz}
\author[c]{Luis Lopes }
\author[a]{Erwin Schwab}
\author[a,b,d]{and Joachim Stroth}

\affiliation[b]{Goethe University Frankfurt, Institute for Nuclear Physics\\Frankfurt, Germany}
\affiliation[c]{LIP, Laboratory of Instrumentation and Experimental Particle Physics, \\Coimbra, Portugal}
\affiliation[a]{GSI Helmholtz Center for Heavy Ion Research,\\Darmstadt, Germany}
\affiliation[d]{Helmholtz Res. Acad. Hesse for FAIR}
\affiliation[1]{corresponding author}
% e-mail addresses: only for the corresponding author
%\emailAdd{c.wendisch@gsi.de}

\begin{abstract}
%% Text of abstract
The central tracking system of the HADES detector, installed at the SIS-18 synchrotron at GSI/Darmstadt (Germany), employs large-area, low-mass drift chambers, featuring Aluminum potential wires and small cell sizes.
The chambers in front of the magnetic field, closest to the interaction point, have developed significant self-sustained currents and discharges during operation, most probably triggered by isobutane-based gas mixtures. 
Only both, (i) replacing isobutane by CO$_2$ and (ii) adding 1000 to 3500 ppmv of water into the Ar/CO$_2$  counting gas mixture, individually optimized for a given chamber, allowed to recover the chambers, enabling stable operation in several production runs since then, e.g. with high-intensity heavy-ion induced reactions. 
The origin of the instability was found to be deposits on the cathode wires, provoking the Malter-like effects, by visual inspection and energy-dispersive X-ray spectroscopy. The charge on the wires accumulated during their lifetime does not point to so-called classical aging, but presumably the interaction of isobutane with materials in the gas flow, residual impurities, and reaction products formed in plasma, e.g., built by discharges.

\end{abstract}

%%Graphical abstract
%\begin{graphicalabstract}
%\includegraphics{grabs}
%\end{graphicalabstract}

%%Research highlights
%\begin{highlights}
%\item Research highlight 1
%\item Research highlight 2
%\end{highlights}

\begin{keyword}
%% keywords here, in the form: keyword \sep keyword
Cathode aging \sep
Wire chamber aging \sep
Polymer growth \sep
Malter effect \sep 
Drift chamber \sep
Water additive \sep
Isobutane

%% PACS codes here, in the form: \PACS code \sep code

%% MSC codes here, in the form: \MSC code \sep code
%% or \MSC[2008] code \sep code (2000 is the default)

\end{keyword}

\end{frontmatter}

%% \linenumbers

%% main text
\section{The HADES drift chambers}
\label{intro}

The High Acceptance Di-Electron Spectrometer HADES \cite{lit:HADES} is a compact high-precision experiment optimized for lepton pair spectroscopy, which is located at the Heavy Ion Synchrotron SIS18 at GSI Helmholtz Centre for Heavy Ion Research, Darmstadt, Germany. It is dedicated to the investigation of relativistic heavy-ion collisions with the goal of studying the properties of QCD matter under extreme conditions. The HADES detector can be operated at interaction rates of up to 20 kHz and features about 100.000 individual detector channels. 

The tracking system is composed of so-called Mini-Drift chambers (MDC), two chamber planes before (I, II), and two behind (III, IV) the toroidal magnetic field, respectively. The 24 drift chambers of the HADES experiment have been designed and built by four different institutes (GSI Darmstadt (I), JINR Dubna (II), HZDR Dresden-Rossendorf (III, I rebuilt) and IPN Orsay (IV)). The full detector is divided into six identical sectors in azimuth, hence equipped with 24 chambers covering 33 m$^2$ active area. Each drift chamber comprises six stereo-angle wire layers. This, together with employing small, i.e.~"mini", drift cells guarantees the granularity and redundancy required by the physics program of HADES. Besides the small cell sizes, the chambers feature a challenging active to inactive geometry to hide the chamber frames, housing the front-end electronics, inside the solid angle defined by the coil boxes of the superconducting magnet. This poses the need for smart mechanical design solutions for the four different chamber types in order to guarantee mechanical stability despite of the forces caused by the accumulated wire tension. The chambers are built of individual wire layers, stacked and sandwiched by dedicated Aluminum window frames. O-rings are introduced in between the layers to allow for reworkability in case of damage to wires, e.g.~tension loss or wire rupture. 

\begin{table}[t]
    \centering
    \begin{tabular}{|l|l|l|l|l|l|}
    \hline
      Parameter   & I   &  I.1 & II  & III & IV \\
    \hline
      Anode pitch (mm)   & 5 & 5 & 6 & 12 & 14 \\
    \hline
      Width (mm)   & 5  & 5 & 5 & 8 & 10\\
    \hline
      $\varnothing$ anode wire ($\mu$m)  & 20 & 20 & 20 & 20 & 30\\
    \hline
      $\varnothing$ F, C wires ($\mu$m)  & 80 & 76 & 80 & 100 & 100$^*$\\
    \hline
      F, C wire material  & Al  & Cu/Be & Al & Al & Al\\
    \hline
      x/X$_0$ (10$^{-4}$), He  & 5.5 & - & 5.2  & 5.0  & 4.8 \\
     \hline
      x/X$_0$ (10$^{-4}$), Ar  & - & 21 & 7.3 & 8.3 & 8.9 \\
   \hline
      Gas flow (vol/h)   & 0.8 & 0.8 & 0.6 & 0.13 & 0.08 \\
    \hline
      HV (-kV), Ar   & 1.75 & 1.75  & 1.77  & 1.9  & 2.15 \\
    \hline
    \end{tabular}
    \caption{Key (operation) parameters for the four different chamber types I-IV. Type I was rebuilt (I.1), see text. Potential wires comprise cathode (C) and (F) field wires. The high voltage (HV) setting at working point is given for Ar:CO$_2$ = 70:30 (Ar). The material budget is also given for the initially used gas mixture He:Isobutane = 60:40 (He). All anode wires are made of W (Au), $^*$: Au-plated.}
    \label{tab:parameters}
\end{table}

High-precision dielectron spectroscopy calls for minimizing multiple scattering.  This is realized by using bare Aluminum potential wires 
for the cathode planes separating and terminating the stereo cell layers, and the field wires separating the adjacent drift cells per layer. In addition, a low-mass gas mixture was initially chosen, He-isobutane gas mixture (60:40), where isobutane acts for both, delivering primary and secondary electrons for ionizing particle tracks, and efficient quencher. Long-chain hydrocarbons such as isobutane are known to trigger aging effects in wire chambers \cite{lit:aging1}. Hence, accelerated aging tests conducted during the prototyping phase demonstrated that so-called "classical" aging due to polymerization of long-chain hydrocarbon radicals resulting from radiation interacting with isobutane is not seen up to 20 mC/cm accumulated charge \cite{lit:MDCaging} of the anode wire. The lifetime accumulated charge of the chambers close to the interaction point is estimated to be close to 15 mC/cm (anode wire) so far, equivalent to approximately 10 (5) mC/cm on the cathode (field) Aluminum wires, and comprising several heavy ion runs, where delta-electrons significantly contribute to the charge load. This, and the fixed-target geometry of the experiment, results in a rather moderate current density range between (below) 1 and 5 nA/cm, with maximum gains for the smallest cell sizes of about 5 $\times$ 10$^5$.
Note, the accumulated charge employing isobutane is significantly below the lifetime numbers stated above due to changing to CO$_2$ forced by operation instabilities as reported in the following, and does not reach the recommended limits for safe operation with hydrocarbon-based gases stated in literature \cite{lit:aging2}.\\
The material budget x/X$_0$  for the four different chamber types is below 0.06 per chamber. 
The gas is recirculated, employing oxygen purifiers, and dedicated means to control the gas quality during run time.  The flow is adjusted to below 1 and 0.13 volume exchanges/hour for the chamber types I\&II and III\&IV, respectively, and is optimized to limit the residual force to the chamber frames, avoiding mechanical deformation and wire tension loss. See table \ref{tab:parameters} for the key parameters of the different chamber types.
 
\section{Observations}
\label{observations}

The HADES drift chambers have been designed and built around the year 2000. The necessity of minimizing multiple scattering in high-precision di-electron spectroscopy called for a He-based gas mixture, with 40~\% isobutane serving both, efficient quencher and providing sufficient primary and secondary electrons to allow for efficient particle detection at a moderate gain of about 10$^5$ at working point. In the following, we report on the observations made with the first (I) and the second (II) MDC plane in front of the magnetic field. 

 \subsection{MDC I: Si deposits}
\label{silicone}
\begin{figure}[tbp]
\centering
\includegraphics[width=.45\textwidth]{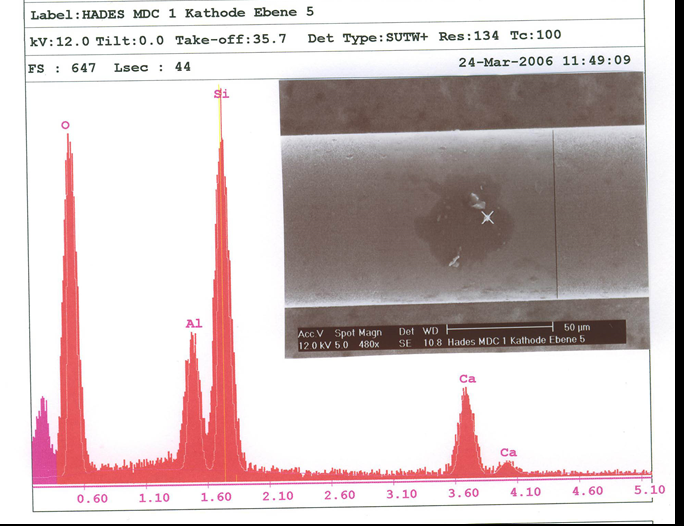}
\caption{\label{fig:si} Energy dispersive x-ray spectroscopy (EDX) of deposits on cathode wire (80µm, Al ) .}
\label{fig:Si}
\end{figure}

Already during the first five years of operation, at that time in experiments with rather low-intensity carbon beam (current load measured on the cathodes $<<$ 1 nA/cm), we observed a slow, but significant increase of localized HV instabilities present in individual wire layers, comprising self-sustained (off-spill) currents, sudden discharges followed by HV trips. All six chambers of this type were affected in about 14 out of 36 wire layers. Any attempts to mitigate the problems with conditioning and increase of gas flow did not result in any improvement.
Opening affected chambers after wire rapture allowed for the inspection of the surface of Aluminum wires. Traces of silicon have been found inspecting deposits on the wires, see figure \ref{fig:Si}. It is well known \cite{lit:Si} that silicon does easily polymerize with hydrocarbons and oxygen, forming heavy polymers not easy to remove.

Changing the gas mixture to a Ar/isobutane mixture, in the hope of stabilizing the operation by generating 
more primary and secondary electrons which allows to reduce the gain,
did not solve the problem. Adding water, as recommended in \cite{lit:additives} triggered an increase of micro discharges not allowing to operate the detectors. This was most probably caused by a mechanical peculiarity of this chamber type, employing stainless steel frames attached to individual wire layers to reinforce mechanical stability, partly exposed to the counting gas. A topological analysis of the occurrence of the observed instabilities in all chambers affected pointed to the definite origin of the silicon, which was an O-ring close to one of the two window frames of a given chamber, obviously accidentally being treated with standard vacuum grease (\textit{Lithelen}\textregistered, produced by Leybold, containing silicon compounds) during fabrication. 
As a consequence, we were forced to rebuild all six (and one spare) chambers of plane I, called I.1 in table \ref{tab:parameters}, which have been successfully set into operation in a heavy ion experiment in 2012. It is important to note that we did not find traces of silicon distributed in the gas system of the MDC tracking system by inspecting gas filters installed in the input line of all chambers. 

\subsubsection{MDC I.1 and II: Malter-like effects}
\label{malter}
\begin{figure}[tbp]
\centering
\includegraphics[width=.45\textwidth]{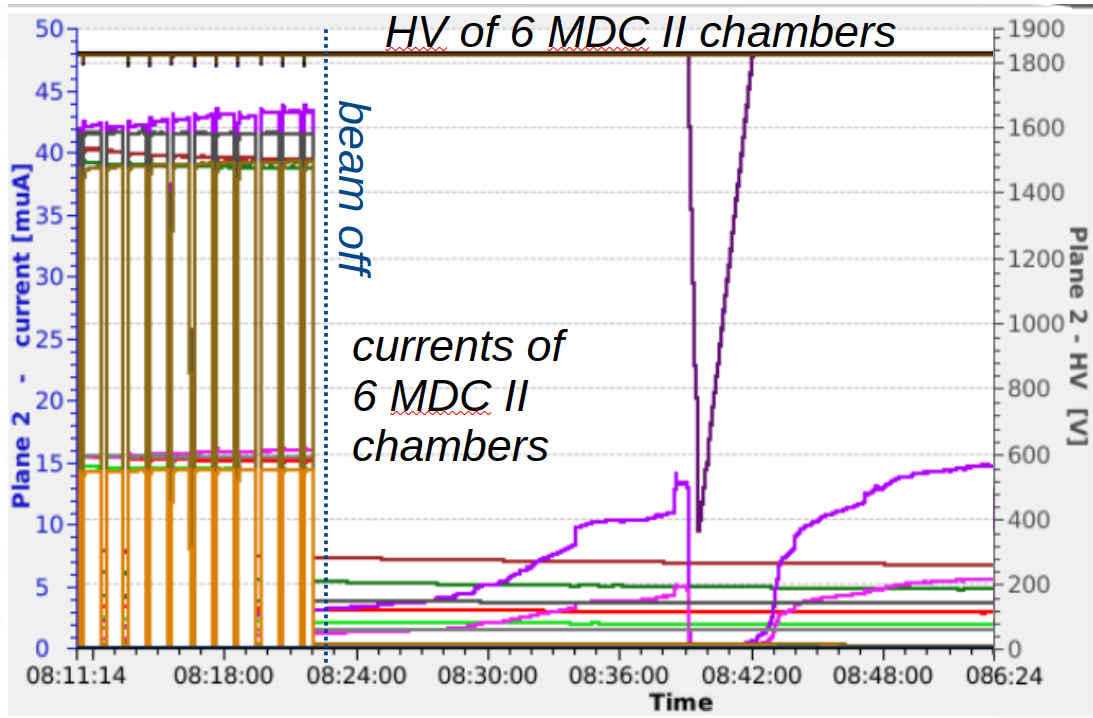}
\caption{\label{fig:selfsustained}
HV (-V) and current ($\mu$A) trends of six MDC II chambers during beam irradiation, exhibiting in- and off-spill self-sustained currents in a given chamber developing with minutes-time scale. HV cycling (triangular HV trend) did not lead to a sustained suppression of the self-sustained currents.}
\end{figure}

Both chamber types I-IV (I.1) have been exposed to isobutane-based counting gases during the first 12 (1) years of operation (commissioning). Type II started to exhibit operation instabilities, mainly self-sustained currents, during a heavy ion run in 2012, with maximum current densities not exceeding several nA/cm. Figure \ref{fig:selfsustained} shows an example of these self-sustained currents in a given chamber (II), both in-spill (adding to the particle-induced currents) and off-spill (in between spills as well as during beam breaks). Note, sometimes (not always) the self-sustained currents re-establish after HV cycle without radiation.

\begin{figure}[tbp]
\centering
\includegraphics[width=.45\textwidth]{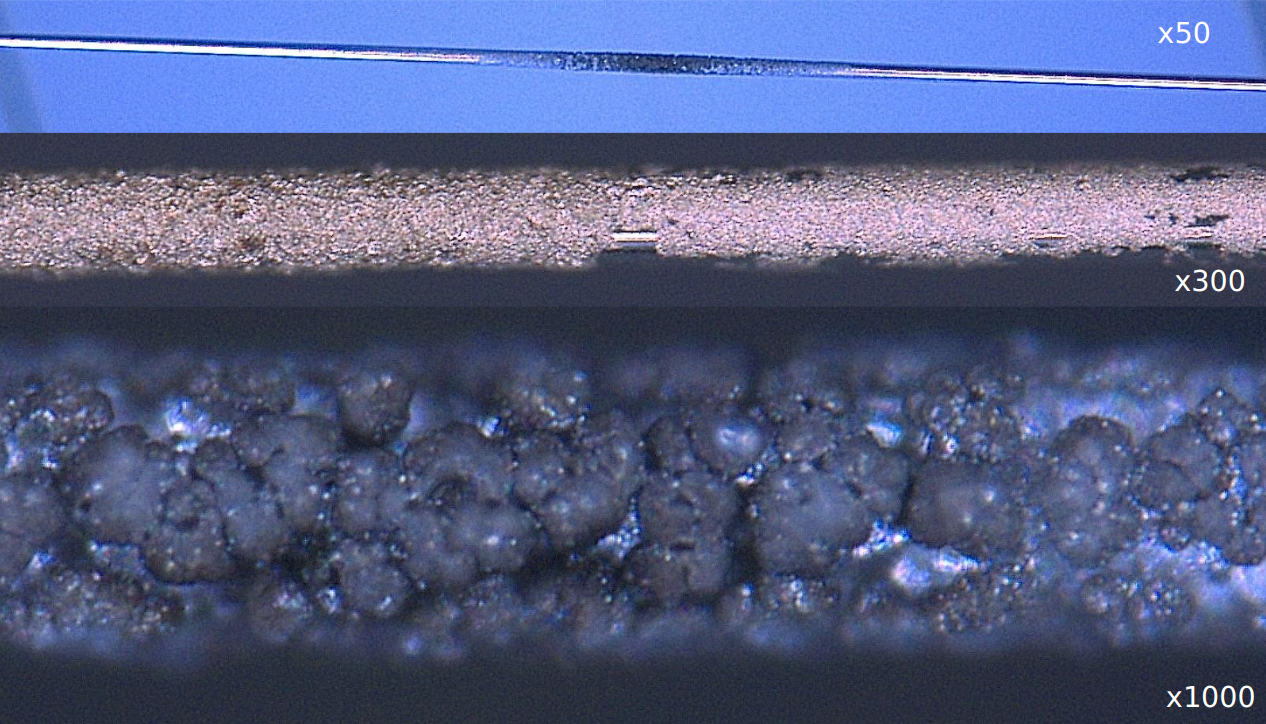}
\caption{\label{fig:darkspots} Visual inspection of cathode wires (Al, 80 µm) MDC II, 2016, exhibiting dark spots. EDX analysis indicates that they are mainly formed by hydrocarbon compounds.}
\end{figure}

Chamber type I.1 did start to exhibit similar instabilities, but less prominent. On the occasion of a wire rupture in a type II chamber, cathode and field wire surfaces have been inspected. Figure \ref{fig:darkspots} displays the results with different zoom levels. The dark spots have been visually found just be eye on all cathode wire planes, homogeneously distributed over the active area with roughly one dark spot per 1 to 2 cm$^2$. EDX analysis points to hydrocarbon compounds, but silicon was not found, which also confirms the observation reported above, that silicon compounds was not migrating in the gas system.\\
As a consequence, isobutane was completely banned from the MDC gas system by 2014, following the recommendations in literature, e.g.~\cite{lit:aging2}. Both, the current densities on the wires during operation as well as the accumulated charge per wire length during lifetime  so far (below 10 mC/cm on the cathode wires) did not suggest "classical" aging causing the observations. However, the observations made pointed to Malter effect on the cathode wires. A loss of gain was not observed.\\
By employing a Ar/CO$_2$-based gas mixture we were forced to trade operation stability with data quality.
The increased electron drift velocity leads to a loss in the precision (time measurement and spatial) of 10 to 15\% in the given reduced electrical field range. In addition, the variation of drift velocity in this range increases by employing CO$_2$, impacting the time-position calibration. A distinctive feature of the rectangular drift cells are the corners, featuring rather low reduced electrical fields. Garfield simulations indicate a significantly lower drift velocity for CO$_2$-based gases compared to isobutane as a quencher in this reduced electrical field range, which leads to a detection efficiency drop in the corners due to insufficient charge collection (this effect is presently being assessed and corrected for in data analysis).

%%% --- CM ----

\section{Method of restoring stable operation}
\label{restoring}
With the observations reported above, mainly for chamber type II during 2012, we have revisited the option to use additives in order to mitigate the instabilities and allow for reliable operation in the upcoming production experiments. 
Following \cite{lit:aging1, lit:aging2,lit:additives}, we tried to identify the optimum operating value w.r.t.~the relative concentration of water in the gas (Ar/CO$_2$ 70/30). Deionized water (and some alcohols) is known (i) to suppress the polymerization of radicals and (ii) reduces charging-up effects on insulating surfaces, such as deposits on wires. Obviously, a careful systematic assessment is mandatory for each chamber type, and even individual chambers, to identify the sweet spot balancing the above-mentioned positive effects and surface conductivity, and intrusion of water in the matrix of e.g.~the frame material (Stesalit\textregistered, Durostone\textregistered). In the following, this procedure as well as the result is presented.

\subsection{Adjusting the water additive and operation}
\label{water}
\begin{figure}[tbp]
\centering
\includegraphics[width=.45\textwidth]{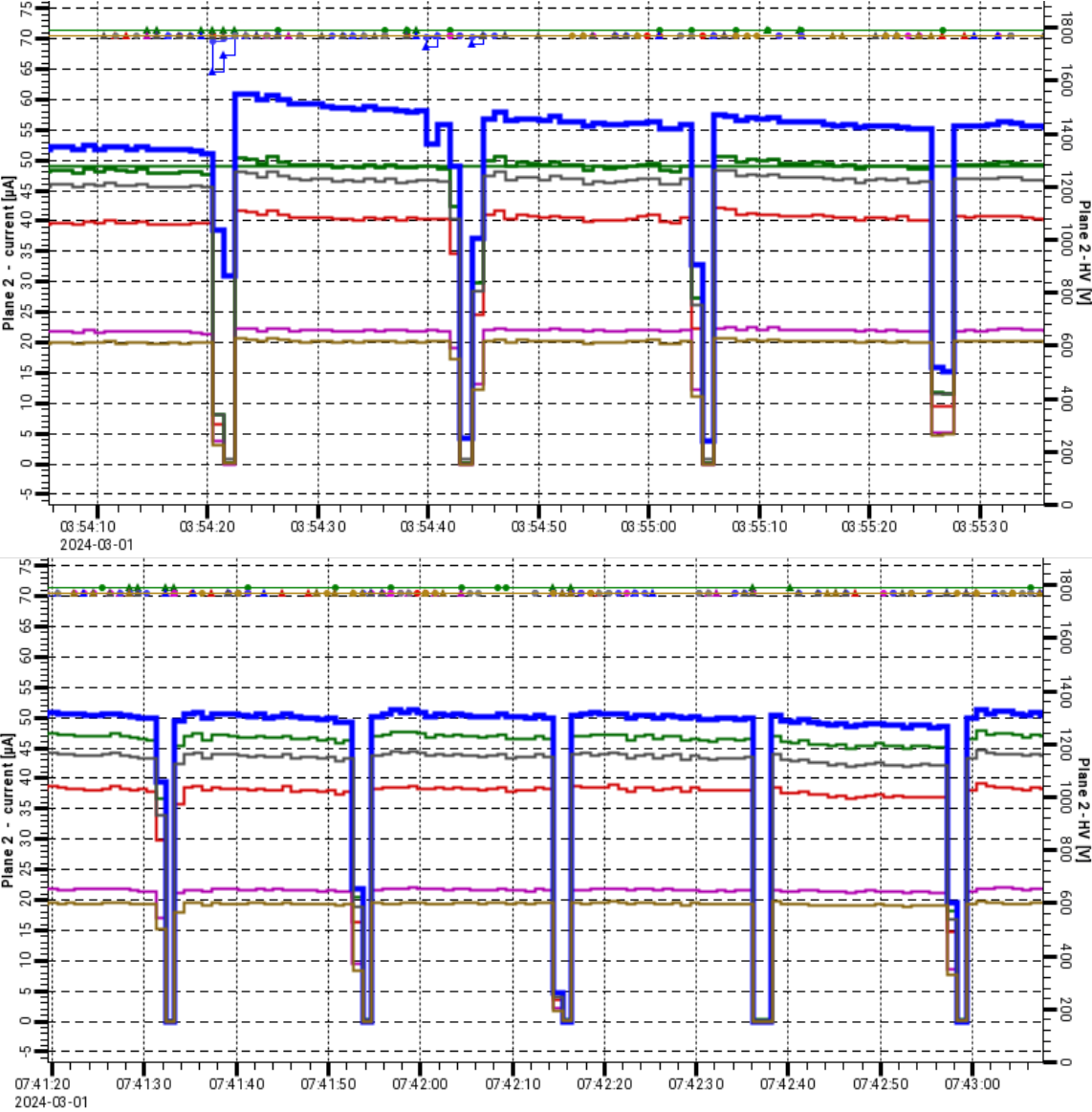 }
\caption{\label{fig:watertrend} HV (-V) and current ($\mu$A) trends of six chambers type II during a heavy ion beam time (Au+Au reactions, $E_{kin}=$ 0.8A GeV), with water vapor additive individually adjusted for each chamber. Upper plot, discharging of a self-sustained current of the chamber in blue. Lower plot, all self-sustained currents resolved. Note, chamber-to-chamber differences of the in-spill DC cathode currents are caused by a different shielding of $\delta$-electrons in front of the given chambers. } 

%drift chambers represented by magenta and brown lines exhibit a factor 2 lower particle load, due to more effective shielding of $\delta$-electrons by glas mirrors (of the RICH detector ) present in this two sectors only, instead of carbon fiber mirrors for the other 4 sectors.   }
\end{figure}

The procedure described here was applied for the 12 chambers in front of the magnetic field which exhibit Malter-like self-sustained current during irradiation, also with Ar/CO$_2$ counting gas. Since the purifiers used to clean the recycled gas would affect the water admixture, the gas supply of these chambers was switched to an open system. Water vapor and oxygen concentration, as well as the over-pressure (30-50 Pa), are being monitored continuously. 
Part of the fresh gas is taken to be enriched with deionized water, employing a temperature-controlled gas washing bottle. Thus, this ("wet") gas line is precisely adjusted to a defined concentration (100\% humidity equals 15730 ppmv at 14°C) of water. Then this wet line is mixed with the dry line of counting gas with mass flow-meters. As a result, chambers, or groups of chambers, can be supplied with an individually adjusted absolute water concentration. With systematic studies prior to physics campaigns and described in the following, coarse individual water concentrations have been determined. Here, we took the profit of irradiating the chambers with a X-ray tube, mimicking the runtime current density (with a safety factor) and its polar dependence due to the fixed target setup of the experiment. Moreover, the X-ray tube is operated with a duty cycle typical for the one delivered by the SIS18 accelerator in a heavy ion experiment. In the ideal case, the chamber under test does not exhibit any measurable residual current during the spill breaks (typically some seconds). These tests with individual chambers usually last several hours or days also used for conditioning prior to runtime.
Figure \ref{fig:watertrend} shows the current over time after having tuned a chamber w.r.t.~water concentration needed for stable operation. The beam spill structure is clearly visible, as well as the absence of any self-sustained currents.\\   

It is necessary to fine-tune the water vapor concentrations during runtime to counteract beginning self-sustained currents, triggered either by radiation (action: increase of water concentration) or surface conductivity (decrease). Typically, currents caused by surface effects set in with a very small value and increase slowly (timescale hours), in contrast to Malter-induced currents Table \ref{tab:water} depicts typical ranges of deionized water concentration after fine-tuning for chambers of type II.  

\begin{table}
    \centering
    \begin{tabular}{|c|c|c|c|c|c|c|} \hline
     Chamber    & 1 & 2 & 3 & 4 & 5 & 6 \\ \hline
     min. & 1800 & 1000 & 2000 & 3000  & 1800  &  2500\\
     (ppmv) & & & & & &  \\ \hline
     max.  & 2100  & 1800 & 3000  & 3500  & 2000  & 3000 \\
     (ppmv) & & & & & & \\ \hline   
    \end{tabular}
    \caption{Typical ranges of individual deionized water concentrations for chambers of type II.}
    \label{tab:water}
\end{table}

It is interesting to note that the mean values and ranges differ from chamber to chamber. Most probably this is caused by the different levels of contamination with deposits and might be closely linked to the individual history of the chambers, both w.r.t.~ manufacturing and material (wire) quality as well as operation history, e.g.~occurrence of discharges. Presently, chambers of type I.1 are being operated with the addition of deionized water vapor, however, with lower concentrations ($<$ 1500 ppmv) compared to type II . Also, chambers can be grouped by being exposed to the same water concentration.

\section{Summary and conclusion}
\label{conlcusion}

This paper reports on operation experiences with low-mass drift chambers over more than two decades of operation in the fixed-target HADES experiment at SIS18, GSI Darmstadt. The low-mass concept called for a He/isobutane counting gas and the use of aluminum for cathode and field wires. The 24 chambers of four different sizes (types) feature small drift cells to provide the granularity requested by the physics program.  
Already during the first years of operation all chambers of type I, the smallest chambers close to the target, exhibited self-sustained currents and sudden discharges during runtime, conditioning did not improve stability. The investigation of visible deposits on cathode wires proved the presence of silicon, which could be traced back to the vacuum grease used for part of the O-rings containing silicone. 
Additives to mitigate self-sustained currents, resulted into significant increase of discharge rate, thus the rebuild of all type I chambers was the only solution.\\
In the second half of the HADES lifetime, it became obvious that the use of isobutane as an efficient quencher gas but also a well-known source of radicals under irradiation and creation inside a plasma, e.g., discharges, can no longer be advised. All chambers in front of the magnetic field exhibited self-sustained currents, followed by discharges and HV trips, if not extinguished by decreasing the gain. It is interesting to note that our observations are not compatible with the so-called "classical" aging of wire chambers operated with isobutane as quencher, since both, the typical current densities of only several nA/cm on the wires, as well as the total accumulated lifetime charge of below 15 mC/cm (anode wire), does not come close to the values reported in dedicated literature 
($>$ 100 mC/cm \cite{lit:Si}). In this case we can only speculate on the reason which has affected the operation stability of the chambers. Materials in contact with the gas (valves, sealing) that react/are dissolved with isobutane cannot be excluded to be a source for deposits on the (cathode) wires. The sandwich concept of the chamber construction, allowing for reworkability, leads to residual gas leakage, with typical oxygen contamination of 100  ppmv in a given chamber. This would in principle allow for contamination of the gas volume with environmental substances. 
%%However, the presence of oxygen and water (environment humidity) could also contribute to stabilize chamber operation. 
A third source for the observed "accelerated" aging might be uncontrolled discharges during operation, due to wire surface defects.  
%which cannot be easily avoided be means of HV monitoring (current limits).
The plasma in a discharge leads to dissociative chemical processes with products affecting all (adjacent) surfaces. Also, discharges themselves might damage the wire surface, which would act as a seed for polymerization.\\
Adding individual, well-adjusted levels of deionized water vapor, i.e., concentrations between 1000 and 3500 ppmv, did significantly stabilized the operation of all affected HADES chambers. It is important to control the amount of water vapor during runtime to always dynamically meet the sweet spot, balancing between surface conductivity (talking about all surfaces in the gas volume) and suppressing self-sustained Malter-like currents by acting on the radicals and increasing surface conductivity of non-conductive films created by polymerization of hydrocarbons. It is important to emphasize that the addition of water does not act on the origin of the self-sustained currents, i.e., removing the water  leads to the previously observed operation instabilities.
%, while restoring the water concentration does restore smooth operation.
This has been demonstrated since then in several, also high-intensity, physics experiments lasting several weeks, and recommends employing the HADES drift chambers also in future experiments at FAIR with larger current densities on the wires tested successfully by the x-ray conditioning.

%% The Appendices part is started with the command \appendix;
%% appendix sections are then done as normal sections
%% \appendix

%% \section{}
%% \label{}

%% If you have bibdatabase file and want bibtex to generate the
%% bibitems, please use
%%
%%  \bibliographystyle{elsarticle-harv} 
%%  \bibliography{<your bibdatabase>}

%% else use the following coding to input the bibitems directly in the
%% TeX file.

\section{Acknowledgments}
This work, covering several funding periods, has been supported by the BMBF and GSI. 
We also like to acknowledge the support of many people from various institutes, GSI Darmstadt, JINR Dubna, HZDR Dresden-Rossendorf and IPN Orsay, participating in the design, construction, and operation of the chambers over more than two decades. The project reported here was over the years accompanied and consulted by J. Pietraszko, L. Naumann, L. Heinrich, J . Hutsch, O. Fateev, and C. Garabatos, who deserve special thanks.

\end{document}